\documentclass[aps,preprint,prd,epsfig]{revtex4}
\usepackage{amssymb}
\usepackage{amsmath}
\usepackage{graphicx}
\usepackage{fancyhdr}
\newcommand{\be}{\begin{equation}}
\newcommand{\ee}{\end{equation}}
\newcommand{\bea}{\begin{eqnarray}}
\newcommand{\eea}{\end{eqnarray}}
\newcommand{\bes}{\begin{subequations}}
\newcommand{\ees}{\end{subequations}}

\newcommand{\bc}{\begin{center}}
\newcommand{\ec}{\end{center}}
\usepackage{amsmath, amssymb} 
\usepackage{slashed}

\begin{document}

\title{ A cosmologically viable eV sterile neutrino model.}

\author{C. A. de S. Pires}
\affiliation{{ Departamento de
F\'{\i}sica, Universidade Federal da Para\'\i ba, Caixa Postal 5008, 58051-970,
Jo\~ao Pessoa, PB, Brasil}}

\date{\today}

\begin{abstract}
The MiniBooNE collaboration recently released a report claiming have observed  an excess of electron and anti-electron neutrino with significance of $4.8 \, \sigma$ C.L. corroborating, in this way, the long-standing   LSND anomaly. Combined LSND and MiniBooNE analysis reach a significance of $6.0\, \sigma$ C.L. Such a result, if confirmed by future experiments, will cause considerable impact on particle physics since that such anomalies, when interpreted  in terms of neutrino oscillation, require the existence of at least one light sterile neutrino. It happens that, on according to standard scenarios,  such light sterile neutrino is incompatible with current cosmological data. In this way, understand these anomalies require an extension of the standard model capable of generating tiny masses for both active and sterile neutrinos and re-conciliates such a result  with cosmology. An interesting proposal in this direction involve the existence of a secret sector interacting exclusively with sterile neutrinos.  In this work we implement the canonical seesaw mechanism into the standard model in such a way that generates tiny masses to the active and sterile neutrinos and  embody a secret sector capable of re-conciliating eV sterile neutrinos with cosmology. As other gains,  the scalar content required by the implementation of the mechanism provides contribution to rare lepton decays, may accommodate the $g-2$ of the muon and poses a scalar singlet that may drive  inflation through   Higgs inflation mechanism without problem with loss of unitarity. 

\end{abstract}

\maketitle

\section{Introduction}

The physics of neutrinos keep being the most active and fascinating branch of particle physics. This is so because we are far from a final understanding of the origin of  neutrino masses and mixing. For example,  neutrino anomalies  are hinting by the existence of at least one sterile neutrino (or right-handed neutrino) with mass at eV  scale and largely mixed with the standard neutrinos\cite{lsnd,reactor,gallium,miniboone}. It happens that, according to standard scenarios, eV sterile neutrinos are strongly disfavoured by current cosmological data\cite{BBN,planck2018,Neff}. In view of this, we think that, if future short-baseline experiments confirm the existence of eV sterile neutrinos\cite{neos,dans}, we, then, have to find a new road to understand the origin of neutrino mass and mixing. This is so because heavy right-handed neutrinos provide the best way for understanding  the origin of the smallness of the standard neutrino masses by means of the canonical type I seesaw mechanism\cite{seesawI}. So if right-handed neutrinos are confirmed to be light, we are forced to re-interpret the canonical type I seesaw mechanism.

On the other hand, if we insist in the explanation of the SBL anomalies in terms of neutrino oscillation, we have to re-conciliate  light right-handed neutrinos with cosmology. It has being recently proposed that such re-conciliation  requires the existence of a secret sector involving  very light degree of freedom in the form of vector boson or pseudo-scalar interacting exclusively with the sterile neutrinos\cite{reconc1,reconc2}. The case of pseudo-scalar is particularly interesting because, as discussed here,  such degree of freedom  may be the remnant of the spontaneous violation of an accidental global symmetry  at very low energy scale.

Recapitulating. On one hand, the canonical seesaw mechanism, as we understand it, requires that  right-handed neutrinos be heavy particles and that  lepton number be violated  at very high energy scale. On the other hand,  the state of the art  in neutrino physics ( taking into account neutrino anomalies)  may be pointing to the contrary: light right-handed  neutrinos with lepton number being violated at very low energy scale. 

 In this work we propose an extension of the standard model which embody the canonical type I seesaw mechanism in such a way that lead to light active and sterile neutrinos and contains a pseudo-scalar carrying the  features necessary to  re-conciliate SBL anomalies with cosmology.  Our model is constrained by rare lepton decay and gives contribution to the $g-2$ of the muon. As gain, by triggering non-minimal coupling of the scalars  with gravity, the model may perform Higgs inflation without face problem with loss of unitarity.

\section{The model}
%
\subsection{eV sterile neutrino and cosmology}

Since the early 2000s  neutrino anomalies have been claiming by the existence of eV sterile neutrinos\cite{lsnd}. Recently  MiniBooNE collaboration released a report  corroborating the previous LSND results\cite{miniboone} which revived the interest  in light sterile neutrinos. These experiments claim have detected an excess of $\bar \nu_e$ and $\nu_e$ events coming from $\pi^+$ decay that when interpreted in terms of neutrino oscillation requires the existence of a fourth neutrino with mass around eV scale largely mixed with the standard ones. In spite of  Gallium and Reactor experiments confirm LSND and MiniBooNE results \cite{reactor,gallium}, the existence of light sterile neutrinos is not definitive,yet.  This is so because there are alternative explanations for these results \cite{alternative} and MINOS+ and ICECUBE experiments shed doubt on the neutrino oscillation interpretation of the LSND and MiniBooNE results\cite{minos,icecube}. However, we think that the indications in favor of eV sterile neutrino is strong enough for we start wondering about what kind of new physics may support such results. For review of this subject, see Ref. \cite{review}.

On the other hand, light sterile neutrino is strongly disfavored by current cosmological data involving big bang nucleosynthesis (BBN), cosmic microwave background(CBM) anisotropies and large scale structure(LSS)\cite{BBN,planck2018,Neff}. This is so because, in face of the large mixing, neutrino oscillation may conduct sterile neutrino to thermal equilibrium with the active neutrino even before neutrinos decouple from the primordial plasma.  This causes considerable  impact on structure formation,  CMB and BBN. A possible solution for this tension is to suppress the production of these neutrinos in the early universe avoiding, in this way, that  they thermalize with the active ones at high temperature\cite{reconc2}.

A recent proposal going in this direction makes use of what has being called  secret interactions.\cite{reconc1,reconc2}. By secret interaction we mean that there must be a particle, vector boson or pseudo-scalar, that interacts exclusively with the sterile neutrinos. The main feature of such particle is that it has to be lighter than the lightest sterile neutrino. In our case  the secret particle will be  a pseudo-scalar and the secret interaction take the form\cite{reconc1}
\begin{equation}
    \sim g_s \bar \nu^C_S\gamma_5 \nu_S I.
    \label{secretinteracvtion}
    \end{equation}

It has been claimed in \cite{reconc1} that if $m_I << m_{\nu_S}$ with  $g_S$ taking values in the range  $10^{-6}-10^{-5}$  we, then,  have  re-conciliation of cosmology with SBL anomalies. This new interaction suppresses dynamically active-sterile neutrino mixing only at very high temperature. This is so because the existence of $I$ creates a medium in which the sterile neutrinos propagate in it in the early universe. Then these neutrinos will  experience a potential that will suppress the mixing among sterile and active neutrinos. For more detail of the mechanism, see original references \cite{reconc1,reconc2}. As far as we know, such hypothesis have not been implemented in the framework of a realistic model that accommodate all ingredients necessary to realize the idea.

In what follow we propose that the canonical type I seesaw mechanism be performed at very low energy scale with the lepton number being explicitly violated in the potential of the model at sub-eV energy scale. We show that in this case the mechanism will generate tiny masses for both  active and sterile neutrinos and recover the desired secret interaction capable of re-conciliating eV sterile neutrinos with cosmological data.

\subsection{The mechanism}
 Canonical seesaw mechanisms are associated to new degrees of freedom belonging to very high energy scale. In the particular case of type I seesaw mechanism, the new degrees of freedom are right-handed neutrinos in the singlet form(sterile neutrinos). On opposite side, neutrino anomalies are hinting that these neutrinos are light particles. In order to complicate even more the situation,  reconciliation of these anomalies with cosmology may be indicating that lepton number is explicitly violated at very low energy scale. This is so because the mechanism capable of re-conciliating SBL anomalies with cosmology requires the existence of a very light pseudo-scalar that may be a remnant of the explicit violation of the lepton number at very low energy scale. What is missing in this discussion is a realistic  model that perform such idea, namely, a model that generate tiny  masses for both active and sterile neutrinos and that re-conciliate light sterile neutrinos with cosmology. In what follow we propose a model, based in a simple extension of the SM, that do the job successfully.

 For this we modify the standard model by allowing that its leptonic sector involves right-handed neutrinos in singlet form,  
\begin{equation}
    L_{i_L}=
\left (
\begin{array}{c}
\nu_i \\
e_i
\end{array}
\right )_L \,\, ,\,\,\,\, e_{i_R} ,\,\,\,\, \nu_{i_R}, 
\label{lepton} 
\end{equation}
where $i=e\,,\mu\,,\,\tau$.

For we have a successful mechanism and a safe model, the scalar sector of the model must accommodate   two Higgs  doublet, $H_1$ and $H_2$, and two Higgs singlet, $\phi$ and $\sigma$,
\begin{eqnarray}
H_1 = \left (
\begin{array}{c}
H^+_1 \\
H^0_1
\end{array}
\right ),\,H_2 = \left (
\begin{array}{c}
H_2^+ \\
H_2^0
\end{array}
\right ),\, \phi \,,\, \mbox{and} \,\,\,\, \sigma
\label{scalarcont} 
\end{eqnarray}

Moreover, in order to avoid  active neutrinos receive masses from the standard Higgs mechanism, we impose that the interactions of the model be symmetric by the  following set of discrete symmetry: $Z_3 \times Z_2$ with the fields transforming in the  following way:
\begin{equation}
 (L_i, l_{i_R},\, \nu_{i_R}, \phi)\rightarrow w(L_i,  l_{i_R}, \nu_{i_R}, \phi)  
\end{equation}
by $Z_3$ symmetry, and transforming in the following way 
\begin{equation}
(H_2, \sigma, \nu_{i_R}) \rightarrow -(H_2,\sigma, \nu_{i_R})
\end{equation}
by $Z_2$ symmetry, where $w_k=e^{2 \pi i \frac{k}{3}}$. 

With this set of symmetry the Yukawa interactions among scalars and leptons involve  the following terms
\begin{equation}
    {\cal L}^Y \supset Y_l \bar L H_1 l_R + Y_D\bar L \tilde H_2 \nu_R +\frac{1}{2}Y_S\bar \nu_R^C\nu_R \phi + H.c.
    \label{yukawaI}
\end{equation}

When   $H_2$ and $\phi$ develop vacuum expectation value (VEV) different from zero, $\langle H_2\rangle= \frac{v_2}{\sqrt{2}}$ and  $\langle \phi \rangle= \frac{v_\phi}{\sqrt{2}}$, the last two Yukawa interactions above provide  mass terms exclusively for the neutrinos
\begin{equation}
    {\cal L}^\nu_{mass} =  M_D \bar \nu_L \nu_R +\frac{1}{2}M_R\bar \nu_R^C\nu_R  + H.c.,
\label{masstermsI}
\end{equation}
where $M_D=\frac{1}{\sqrt{2}}Y_D v_2$ and   $M_R=\frac{1}{\sqrt{2}}Y_Sv_\phi$. We can rearrange things and, considering as basis $\nu =(\nu_L \,,\, \nu^C_R)^T$, we can write
\begin{equation}
    {\cal L }^\nu_{mass} = \frac{1}{2}\bar \nu^C M^{M+D}\nu + H.c.,
\end{equation}
where
\begin{equation}
 M^{D+M} = \left(\begin{array}{cc}
0 & M_{D}^{T}        \\
\newline               \\
M_{D} & M_{R}
\end{array}\right).
\label{M+D}
\end{equation}
This is the canonical mass matrix of the seesaw mechanism.  When $M_R>> M_D$ we have that block  diagonalization of  $M^{D+M}$ leads to the diagonal block matrix

\begin{equation}
M_{DIAG} = \left(\begin{array}{cc}
M_{light} & 0        \\
\newline               \\
0 & M_{heavy}
\end{array}\right),
\label{block}
\end{equation}
with $M_{light}=\simeq -M_D^T M_R^{-1}M_D$ and $M_{heavy} \simeq M_R$. In terms of the VEVs we have 
\begin{equation}
    M_{light} \simeq - \frac{1}{\sqrt{2}}Y_D^T Y_S^{-1}Y_D \frac{v^2_2}{v_\phi},\,\,\,\,\,\, M_{heavy} =\frac{1}{\sqrt{2}} Y_S v_\phi
    \label{massscale}
\end{equation}
The condition $v_\phi >> v_2$ is inherent of the seesaw mechanism. However the scale of $v_\phi$ and $v_2$ are free to take any value since obey $v_\phi >> v_2$ . In the canonical case, $v_2$ belongs to the weak scale and $v_\phi$ to GUT one. But it is possible to have seesaw mechanism with $v_2$ and $v_\phi$ belonging to other scales. It is this possibility that we will explore here. For example. If we take $v_2=10^2$ eV and $v_\phi=10^5$ eV, we then have $\frac{v_2^2}{v_\phi}=0.1$  eV which fall in the sub-eV scale as required by solar and atmospheric neutrino oscillations.

In the canonical case the reason behind  the choice of $v_2$ and $v_\phi$ is purely theoretical, namely, connect weak scale with grand unified theories. In our case, as developed below, the reason for taking $v_2=10^2$ eV and $v_\phi=10^5$ eV is purely phenomenological, namely, generate sterile neutrinos at ev scale and  re-conciliate SBL anomalies with cosmology.

In what follow we assume that $v_2$ belongs to the eV scale and $v_\phi$ to the keV one. Moreover, for sake of simplicity, we assume $M_{heavy}$ diagonal and  $Y_D$ symmetric. In this case the $6 \times 6$ mixing matrix that diagonalize $M^{D+M}$,  and leads to the physical neutrinos, is given by
\begin{equation}
U \simeq \left(\begin{array}{cc}
U_{PMNS} & (M_R^{-1} M_D)^{\dagger}        \\
\newline               \\
-M^{-1}_R M_D & 1
\end{array}\right).
\label{mixingmatriz}
\end{equation}

As we sad above, our main hypothesis here is that the canonical seesaw mechanism  works for $M_D$ and $M_R$ belonging to  low energy scale. However, the condition to re-conciliate eV sterile neutrino with cosmology is decisive to infer the relative values among $M_D$ and $M_R$. Unfortunately we do not poses constraint necessary to fix all the parameters that compose $M_D$ and $M_R$. Thus, we have to resort to illustrative points. If we find at least one set of points that concretizes the hypothesis, we say it works. As  illustrative benchmark points, which leads to light active and sterile neutrinos and conciliate SBL anomalies with cosmology, we choose
\begin{eqnarray}
&&v_D=10 \,\mbox{eV},\,\, v_\phi=10^5 \,\mbox{eV},\nonumber \\
&&Y_{D_{11}}=0.018, \,\, Y_{D_{22}}=-1.3013,\,\, Y_{D_{33}}=0.3639,\nonumber \\
&&Y_{D_{12}}=-0.0113,\,\, Y_{D_{13}}=-0.0383,\,\, Y_{D_{23}}=-0.7631,\nonumber \\
&&Y_{S_{11}}=2\times10^{-5},\,\, Y_{S_{22}}=0.01, \,\, Y_{S_{33}}=0.1
\end{eqnarray}

This set of  benchmark points, when substituted in $M^{M+D}$ given above,  and after its diagonalization,   provides
\begin{eqnarray}
&& m_{\nu_1} \simeq 2\times 10^{-4}\mbox{eV}\,,\,\,\,m_{\nu_2} \simeq 8,6\times 10^{-3}\mbox{eV}\,,\,\,\,m_{\nu_3} \simeq 5\times 10^{-2}\mbox{eV},\nonumber\\
&&\,\,\,\,\,\,\,\,\,\,\,\,\,\,\, m_{\nu_4} \simeq 1.4\mbox{eV}\,,\,\,\, m_{\nu_5} \simeq 0.7\mbox{keV}\,,\,\,\,m_{\nu_6} \simeq 7\mbox{keV}
\label{maspredictions}
\end{eqnarray}
Observe that the masses of the first three neutrinos  accommodate solar and atmospheric neutrino oscillation experiments, while the mass of the  fourth neutrino fall in the range of value required by SBL anomalies. Perceive that $\nu_5$ and $\nu_6$ are much heavier than $\nu_4$.  This choice recovers the 3+1 scenario.

The neutrino mixing matrix  $U$ that diagonalize $M^{D+M}$ is given by

\begin{equation}
U = \left(\begin{array}{cccccc}
0.83 & 0.54 & -0.12 & 0.045& \sim 10^{-5} & \sim 10^{-6}      \\
-0.25 & 0.59 & 0.72 & -0.030 & -6\times 10^{-3} &\sim 10^{-5}      \\
044 & -0.6 & 0.69 & -0.09 & \sim 10^{-4} & \sim 10^{-5}      \\
-0.045 & 0.03 & 0.09 & 1&  0 & 0     \\
\sim 10^{-6}& \sim 10^{-4}& \sim 10^{-4} & 0&  1 & \sim 0      \\
\sim 10^{-6} & \sim 10^{-5} &- \sim 10^{-4} &0 & 0& 1
\end{array}\right).
\label{mixing6x6}
\end{equation}
See that the first $3\times 3$ block  recovers the PMNS mixing matrix. Moreover, this mixing matrix says that the sterile neutrinos $\nu_5$ and $\nu_6$ decouple from the rest of the neutrinos recovering the $3+1$ scenario. 

 In the 3+1 case  what matter is $m_{\nu_4}$ and the angles $U^2_{e4}$ and $U^2_{\mu4}$. Our benchmark points provides $\nu_4$ with mass around eV scale largely mixed with the standard neutrinos where $U^2_{e4}$\,, $U^2_{\mu4}\sim 10^{-2} - 10^{-3}$.  Thus, the canonical seesaw mechanism working at low energy scale is efficient in generating tiny active and sterile neutrino masses.  accommodating, in this way,  the current SBL anomalies. Next step is to check if such scenario is safe in regarding  cosmological constraints. This will be done in the next section. 

\subsection{Re-conciliation with cosmology}

As developed in the previous section, light neutrinos(active and sterile) may be  achieved by means of the type I seesaw mechanism working in low energy scale. Here we argue that our model allows  conciliation of eV sterile neutrinos with cosmology. For this we show that the model has  a secret sector that interacts exclusively with the sterile neutrinos and that such sector is composed by a  pseudo-scalar with mass at sub-eV scale. This light pseudo-scalar  will be  the remnant of the explicitly breaking of the lepton number that we assume to occur at very low energy scale. This energy scale will be dictate by the condition of re-conciliation of cosmology with eV sterile neutrino. For check this, we have to develop the potential of the model.

With the two doublets, $H_1$ and $H_2$ and the  two singlet $\phi$ and $\sigma$, the terms of the potential that obey the set of discrete symmetry $Z_3 \times Z_2$ are these ones
\begin{eqnarray}
&& V(H_1,H_2,\phi)= \mu_1^2 H_1^{\dagger}H_1 +\mu_2^2 H^{\dagger}_2H_2 + \mu_3^2 \phi^* \phi + \mu_4^2 \sigma^* \sigma + \frac{\lambda_1}{2}(H_1^{\dagger}H_1 )^2 +\nonumber \\
&&+\frac{\lambda_2}{2}(H_2^{\dagger}H_2 )^2 +\frac{\lambda_3}{2}(\phi^* \phi )^2 +\frac{\lambda_4}{2}(\sigma^* \sigma )^2 -\lambda_5 H_1^{\dagger}H_1 H^{\dagger}_2H_2 -\lambda_6 H_1^{\dagger}H_2 H^{\dagger}_2 H_1-\nonumber \\
&&+\frac{\lambda_7}{2}((H_1^{\dagger}H_2)^2 + (H_2^{\dagger}H_1)^2) -\lambda_8 H_1^{\dagger} H_1 \phi^* \phi -\lambda_9 H_1^{\dagger}H_1 \sigma^*\sigma -\lambda_{10} H_2^{\dagger} H_2 \phi^* \phi -\nonumber \\
&&+\lambda_{11} H_2^{\dagger}H_2 \sigma^*\sigma+ \lambda_{12}\sigma^* \sigma \phi^* \phi-\frac{f}{\sqrt{2}}(H_1^{\dagger} H_2 \sigma +H_2^{\dagger}H_1 \sigma^*)-\frac{M}{\sqrt{2}}(\phi^3 +\phi^{*3}).
\label{potential}
\end{eqnarray}
Observe that this potential has two trilinear term which are associated to the energy scale parameters $f$ and $M$. The singlet $\phi$ carries two units of lepton number. Then the last term in the potential violate explicitly lepton number and the parameter $M$ characterizes the energy scale where  this happens.

First step here is expand the fields around their VEVs
\begin{eqnarray}
 H_1^0 , H_2^0 , \phi , \sigma  \rightarrow  \frac{1}{\sqrt{2}} (v_{1 ,2 ,\phi , \sigma} 
+R_{ 1 ,2 ,\phi , \sigma} +iI_{1, 2 ,  \phi , \sigma}),
\label{vacuaII} 
\end{eqnarray}

and then obtain the minimum conditions that guarantee the potential above develop minimum,
\begin{eqnarray}
&&\mu^2_1 -\frac{1}{2}(-\lambda_1 V_1^2+ \lambda_{567}v^2_2 +\lambda_8 v^2_\phi +\lambda_9 v^2_\sigma)-\frac{fv_2 v_\sigma}{2v_1}=0,\nonumber \\
&&\mu^2_2 -\frac{1}{2}(-\lambda_2 V_2^2+ \lambda_{567}v^2_1 +\lambda_{10} v^2_\phi +\lambda_{11} v^2_\sigma)-\frac{fv_1 v_\sigma}{2v_2}=0,\nonumber \\
&&\mu^2_3-\frac{3}{2}Mv_\phi-\frac{1}{2}(-\lambda_3 v^2_\phi +\lambda_8 v^2_1+\lambda_{10}v^2_2)=0,\nonumber \\
&& \mu^2_4 -\frac{1}{2}(-\lambda_4 V_\sigma^2+ \lambda_{11}v^2_2  +\lambda_9 v^2_1)-\frac{fv_2 v_1}{2v_\sigma}=0.
\label{minimum}
\end{eqnarray}
With this in hand, our next step is to obtain  the mass matrices of the scalars. Firstly, let us obtain  the mass matrix of the pseudo-scalars. 

Taking as basis  $(I_1\,,\,I_2\,,\,I_\phi\,,\,I_\sigma)$, the potential above provides the following mass matrix for the pseudo-scalars of the model
\begin{equation}
M_I^2 = \left(\begin{array}{cccc}
\frac{(fv_\sigma+2\lambda_7 v_1 v_2)v_2}{4v_1} & -\frac{1}{4}(fv_\sigma+2\lambda_7 v_1 v_2) & 0 & -\frac{fv_2}{4} \\
-\frac{1}{4}(fv_\sigma+2\lambda_7 v_1 v_2) & \frac{v_1}{4v_2}(fv_\sigma+2\lambda_7 v_1 v_2) & 0 & \frac{fv_1}{4} \\
0 & 0 & \frac{9}{4}Mv_\phi & 0 \\
-\frac{fv_2}{4}  & \frac{fv_1}{4} & 0 & \frac{fv_1v_2}{4v_\sigma}
\end{array}\right).
\label{massmatrixODD}
\end{equation}
Analyzing this mass matrix, we see that the pseudo-scalar $I_\phi$ decouples from the other ones. Consequently it couples exclusively with the right-handed neutrinos(sterile neutrinos). The secret interaction arises from the  last term in Eq. (\ref{yukawaI})
\begin{equation}
    \frac{i}{2}Y_S\bar \nu_R^C I_\phi \nu_R  + H.c.
    \label{secretinteraction}
\end{equation}
This realizes the secret interaction  required to re-conciliate eV sterile neutrino with cosmology, see  Eq. (\ref{secretinteracvtion}). The second condition is that such pseudo-scalar has mass at sub-eV scale. From the mass matrix above, we obtain the following   expression to its mass,
\begin{equation}
  M^2_{I_\phi}=\frac{9}{4}M v_\phi.
  \label{massPS}
\end{equation}
Following  Refs. \cite{reconc1,reconc2}, which suggest  $M_{I_\phi}<0.1$ eV, we then obtain the condition $M< 10^{-8}$ eV for $v_\phi=10^5$ eV. In other words, re-conciliaton of eV sterile neutrino with cosmology, according to our scenario,  is requiring that lepton number be violated explicitly at very low energy scale. This is a very original result. 

Moreover, observe  that our benchmark points above provide one sterile neutrino, $\nu_6$, with mass of 7 keV. In this case,  we have that  $\nu_6$  may be  dark matter in the form of warm dark matter\cite{kusenko}. Although $\nu_6$ is not stable, because $Z_2$ is broken when $\sigma$ develop VEV, its  lifetime is much longer than the age of the universe. This is so because its mixing with the standard neutrino will induce  the radiative  decay  $\nu_6 \rightarrow \nu_e \gamma $ which is the only way such neutrino may decay. The   width for this decay channel take the expression\cite{decayneutrino},
\begin{equation}
 \Gamma(\nu_6 \rightarrow \nu_e \gamma) =\frac{9}{256 \pi^2}\alpha_{EM}G_F^2 U^2_{e 6}m_{\nu_6 }^5 =\frac{1}{1.8 \times 10^{21} s} U^2_{e6}( \frac{m_{\nu_6 }}{\mbox{keV}} )^5 .
 \label{width}
 \end{equation}
 For the values of $m_{\nu_6}$ and $U_{e6}$ given above, we obtain $\tau_{\nu_6}\sim 10^{28}$s. This lifetime is many order longer than the age of the universe.
 
 The condition for alleviating the tension among eV sterile neutrino and LSS of the universe is that the  pseudo scalar $I_\phi$  interact with the dark matter\cite{reconc2}. This is realized in our model since that   $\nu_6$ interacts with $I_\phi$ according to Eq. (\ref{secretinteraction}). 

In summary, the model we are developing here generate small masses for active and sterile neutrino through a kind of type I seesaw mechanism working in low energy scale. A remnant of the explicit breaking of lepton number in the form of pseudo scalar gain mass at sub-eV scale and than re-conciliate eV sterile neutrino with cosmology through secret interaction as claimed in Refs. \cite{reconc1,reconc2}. In what follow we discuss phenomenological consequences and other gains of the model.

\subsection{Some phenomenological aspects and other gains}

Let us follow with the development of the scalar sector. The other three pseudo-scalars provided by the diagonalization of $M^2_I$ are:   one Goldstone boson, $ A_1=I_1+\frac{v_2}{v_1}I_2$, eaten by the standard neutral gauge boson $Z^0$, and other two massive pseudo-scalars that are a composition of the states $(I_1\,\,,\,\,I_2\,\,,\,\,I_\sigma)$. Observe that if we take $M=0$, then $I_\phi$ becomes a Goldstone. This means that $f$ is not associated to the explicit violation of the lepton number.

Concerning the CP-even scalars, considering the basis $(R_1\,,\,R_2\,,\,R_\phi\,,\,R_\sigma)$ and taking $f > v_\sigma > v_1>>v_\phi>>v_2>>M$, we obtain
\begin{equation}
M_R^2 \sim \left(\begin{array}{cccc}
\frac{1}{2}\lambda_1v^2_1 & -\frac{1}{4}fv_\sigma & -\frac{1}{2}\lambda_8 v_1 v_\phi & -\frac{1}{2}\lambda_9 v_1 v_\sigma \\
-\frac{1}{4}fv_\sigma & \frac{fv_1v_\phi}{4v_2} & -\frac{1}{2}\lambda_{10}v_2 v_\phi & \frac{1}{4}fv_1 \\
 -\frac{1}{2}\lambda_8 v_1 v_\phi  &  -\frac{1}{2}\lambda_{10}v_2 v_\phi& \frac{1}{2}\lambda_3 v^2_\phi  & \frac{\lambda_{12}}{2}v_\phi v_\sigma \\
 -\frac{1}{2}\lambda_9 v_1 v_\sigma  & \frac{1}{4}fv_1 & \frac{\lambda_{12}}{2}v_\phi v_\sigma & \frac{1}{4}\frac{fv_1 v_2}{v_\sigma} +\frac{1}{2}\lambda_4 v^2_\sigma
\end{array}\right).
\label{massmatrixEVEN}
\end{equation}
The  diagonalization of this mass matrix  must provide the standard Higgs  and other three  neutral scalars. No one entry of this mass matrix is negligible. Then all these new neutral scalars mix among themselves. See also that the  $3 \times  3$ entry involves $v_\phi$ which is assumed to be small. In this way it is quite reasonable to expect that one neutral scalar, dominantly singlet, be light.  Consequently the model may give a reasonable contribution to the  $g-2$ of the muon by means of this new neutral scalar. We check this in a future paper.

Concerning the charged scalars, taking as basis $(H^+_1\,,\,H^+_2)$, we have 
\begin{equation}
M_C^2 = \left(\begin{array}{cc}
\frac{1}{2}\frac{v_2}{v_1}(fv_\sigma +(\lambda_6 + \lambda_7)v_1 v_2) & -\frac{1}{2}(fv_\sigma+(\lambda_6+\lambda_7)v_1v_2) \\
-\frac{1}{2}(fv_\sigma+(\lambda_6+\lambda_7)v_1v_2) & \frac{1}{2}\frac{v_1}{v_2}(fv_\sigma+(\lambda_6+\lambda_7)v_1v_2)
\end{array}\right).
\label{massmatrixCC}
\end{equation}
As eigenvalues we obtain one Goldstone $h_1^+=H_1^+ +\frac{v_2}{v_1}H_2^+$ and one eigenstate, $h_2^+=-\frac{v_2}{v_1}H^+_1 +H^+_2$, with mass given by 
$m^2_{h^+_2}\sim \frac{1}{2}fv_\sigma$.
This charged scalar intermediates rare lepton decay as for example $\mu \rightarrow e\gamma$. In this case we have\cite{raredecay}
\begin{equation}
    BR(\mu \rightarrow e\gamma)=\frac{0.3 \alpha(Y_{D_{11}}Y_{D_{12}}+Y_{D_{12}}Y_{D_{22}}+Y_{D_{13}}Y_{D_{33}})^2}{64 \pi G^2_F m^4_{h^{+4}_2}}.
    \label{BR}
\end{equation}
The current upper bound on this decay is $BR(\mu \rightarrow e\gamma) \leq 5.7 \times 10^{-13}$\cite{meg}. This upper bound  will put a constraint on the mass of $h^+_2$ and consequently over $f$ and $v_\sigma$. For the set of values of the Yukawa couplings $Y_D$ of our illustrative example  we obtain the constraint  $fv_\sigma \geq 36 \times 10^6$ GeV$^2$.

In what concern $\sigma$, we added it with the proposal of raising the mass of $R_2$. This is so because in the absence of $\sigma$ the mass of $R_2$ get proportional to $v_2$, as in the model in Ref. \cite{neutrinophilic}. However, if we allow a non-minimal coupling among $\sigma$ and gravity, $\xi \sigma^* \sigma R$, its quartic term, $\frac{ \lambda_4 }{2}(\sigma^* \sigma)^2$,  may drive Higgs inflation\cite{higgsinflation} in the  mode of Ref. \cite{pires} which allows $\xi<1$. This is so because the potential above contains a trilinear term involving $\sigma$. This means that its mass term is not dominated by $\lambda_4$. Instead it depends on $f$, $v_\sigma$ and $v_2$. This is seen in the  $4\times4$  entry of $M^2_R$.  Then, even if we have a very small $\lambda_4 $, as required by the cosmological perturbation parameter $\Delta^2_R \sim 10^{-9}$, the mass of the inflaton $R_\sigma$ may be heavy due to the contribution of the term $\sim f\frac{v_1 v_2}{v_\sigma}$  and then engender reheating easily. 

\section{Conclusion}
In this work we proposed and developed  a simple extension of the standard model that is capable of  explaining neutrino anomalies through neutrino oscillation and is in accordance with cosmology.  For achieve this we extended the electroweak symmetry to embed the  $Z_2\times Z_3$ symmetry and extended the particle content of the standard model to embody  three right-handed neutrinos in singlet form plus one new Higgs doublet and two scalar singlets. The discrete symmetries are necessary to guarantee that neutrino masses are generated exclusively by the new scalar content. When the $Z_3 \times Z_2$ is spontaneously broken by the  VEVs of the additional scalar content, the type I seesaw mechanism is triggered. By assuming that the VEVs involved in the mechanism belong to the low energy scale, we obtain a seesaw mechanism that generates tiny masses for both active and sterile neutrinos.

Re-conciliation of eV sterile neutrino with cosmology is achieved by means of secret interactions involving the sterile neutrinos and a pseudo-scalar  with mass at sub-eV scale.  Such exigency imply that lepton number be violated explicitly in an energy scale of order of $10^{-8}$ eV. 

Thus, we are in facing of a change of paradigm in what concern seesaw mechanism. This is so because canonical seesaw mechanism involves very high energy scale. However, explanation of neutrino anomalies by means of neutrino oscillation in accordance with cosmology is pointing to the contrary: seesaw mechanism being realized in very low energy scale.

Moreover, our model poses new degrees of freedom in the form of neutral and charged scalars that may give reasonable contribution to the g-2 of the muon and  mediate rare lepton decay. 

In addition, the second singlet of scalars, the one that is not involved in the seesaw mechanism,   may play the role of the inflaton in a scenario of Higgs inflation where the non-minimal coupling may be smaller than one which guarantee that the model does not face problem with unitarity.

\begin{acknowledgments}
C.A.S.P  was supported by the CNPq research grants No. 304423/2017-3. 
\end{acknowledgments}

\end{document}